\documentclass{article}

\usepackage[T1]{fontenc}
\usepackage[latin9]{inputenc}
\usepackage{float}
\usepackage{textcomp}
\usepackage{amsmath}
\usepackage{amssymb}
\usepackage{graphicx}
\usepackage{esint}
\usepackage{lmodern}
\usepackage{lscape}
\usepackage{hyperref}
\usepackage[lmodern]{mathdesign}

\usepackage[T1]{fontenc}
\usepackage[english]{babel}
\usepackage[gen]{eurosym}
\usepackage{natbib}

\makeatletter

\providecommand{\tabularnewline}{\\}
\makeatother

\begin{document}
\emph{
\begin{center}
This is an archived copy of an accepted manuscript published in Energy Policy available at \url{http://dx.doi.org/10.1016/j.enpol.2017.06.067}. This manuscript version is made available under the CC-BY-NC-ND 4.0. 
\end{center}
}

\title{When Do Households Invest in Solar Photovoltaics? An Application
of Prospect Theory}
\date{}
\author{Martin Klein\footnote{m.klein@dlr.de}, Marc Deissenroth}
\maketitle
\begin{abstract}
While investments in renewable energy sources (RES) are incentivized
around the world, the policy tools that do so are still poorly understood,
leading to costly misadjustments in many cases. As a case study, the
deployment dynamics of residential solar photovoltaics (PV) invoked
by the German feed-in tariff legislation are investigated. Here we
report a model showing that the question of when people invest in
residential PV systems is found to be not only determined by profitability,
but also by profitability\textquoteright{}s change compared to the
status quo. This finding is interpreted in the light of loss aversion,
a concept developed in Kahneman and Tversky\textquoteright{}s Prospect
Theory. The model is able to reproduce most of the dynamics of the uptake
with only a few financial and behavioral assumptions.\end{abstract}

\section{Introduction}

The majority of countries has RES targets and support policies in
place \citep{REN212016}. Such deployment policies, i.e. the desired
diffusion of RES into the market via remunerations like feed-in tariffs,
tenders or market premiums, can be effective tools in creating a market
pull which fosters the uptake of renewables and can, if well designed,
invoke technological evolution and innovation \citep{Hoppmann2013}.
There is little insight, however, on how to set adequate remuneration
levels and when to adjust them, mainly because the drivers and dynamics
of investment are poorly quantified. The policy instruments that try
to incentivize RES deployment therefore often fail to reach desired
quantities. Costly misadjustments could be avoided with a better 
understanding of deployment and diffusion dynamics.

The modeling of market diffusion of RES and in particular photovoltaics (PV) has attracted a considerable
amount of research interest in recent years. 
While there is a fairly large body of literature
on how to set optimal levels of remunerations via real option analysis (for
an overview see e.g. \citet{Zhang2016}), or how firms would ideally time and size investments under regulatory uncertainty (see e.g. \citet{Chronopoulos2016}), 
a growing body of research shows that the residential sector behaves rather differently. 
For instance, the intention formation of home-owners to adopt PV does not solely depend 
on optimality principles and financial factors (see e.g. \citep{Korcaj2015}). 
Energy policy can benefit from a more detailed consideration of behavior \citep{Allcott2010}.
However, methods that take into account more realistic or boundedly rational decision rules have had little impact on the evaluation of residential deployment dynamics -- 
modeling of small scale investments in RES is challenging since many heterogeneous actors and
motives are involved. 

So far, scholars have focused on the socio-demographics
of home-owners and the evaluation of local peer effects (see e.g. \citep{Bollinger2012,Kwan2012,Rode2016}).
They find that localized peer-to-peer communications reduce barriers
to PV adoption \citep{Rai2013}. Most recently, elaborate agent-based
modeling approaches have been presented by \citet{Palmer2015} and
\citet{Rai2015}, which combine both socio-economic characteristics
and peer effects. While all of these approaches provide a detailed
view of the drivers and boundaries of RES uptake, these evaluations
are relatively hard to trace back and generalize, as they require
granular spatial socio-economic data in the former and relatively
specific agent specification in the latter case. They are therefore
hard to apply to other cases and not reducible to analytic demand
formulas and hence of limited use if to be applied in a whole systems
energy modeling context.

Curve fitting approaches try to fill this gap and relate the economic
profitability of a representative PV project with observed aggregated
deployment rates. \citet{Grau2014} mapped the profitability of PV
onto the deployment observed in Germany via a logarithmic fit function.
A dynamic time lag between investment decision and installation is
proposed, which is reduced in situations when remuneration reductions
are announced. Similar exponential curve fitting contributions have
been made by \citet{Benthem2008} and \citet{Wand2011}, additionally
with technology diffusion terms. Similarly, \citet{Lobel2011} are
applying a logit demand function, where the utility of adoption mainly
depends on profitability and the logarithm of cumulative installations.
All of these approaches, however, provide only limited insight into
the dynamics of observed deployment rates, as they either only focus
on yearly installation values \citep{Benthem2008,Lobel2011,Wand2011}
or must be recalibrated over time to make up for unknown dynamic changes
in the adoption behavior \citep{Grau2014}. Finally, \citet{Leepa2013}
present a time-series analysis of the effect of remuneration cuts
on the investment behavior of PV in Germany and find that step-wise
adjustments temporarily accelerate installments. However, a limit
of their study is that they cannot establish causal relationships.

To summarize, a need for dynamic, fundamental, parsimonious models
which are able to depict the magnitude of PV deployment over time
is identified. The aim of this study is to address this research gap. 

The remainder is structured as follows: Section \ref{sec:Case}
introduces the research case - residential PV deployment in Germany
over the years of 2006-2014 - and explains why this is a useful example to study deployment
dynamics and the interaction with the policy regime.
Section \ref{sec:Methods} is concerned with the methodology, i.e.
the techno-economic modeling of PV systems. A way to calculate mean
internal rates of return via a Monte Carlo simulation method is presented.
The deployment modeling via utilities, and most notably, our proposed
extension with the value function of prospect theory, is presented.
Section \ref{sec:Results} presents a deployment analysis on absolute
level of PV profitability, and most notably, shows how this approach
fails to capture the subyearly investment dynamics. The evaluation
then presents how prospect theory can be used to explain the stylized
features of the subyearly deployment dynamics substantially better.
Section \ref{sec:Discussion} discusses the findings, and points out
to possible shortcomings and extensions of the study. Section \ref{sec:Conclusion} concludes
with policy recommendations.

\section{\label{sec:Case}Research Case - Residential Photovoltaics in Germany}

To study the market diffusion of RES, the case of residential PV deployment in Germany over the years of 2006-2014 is investigated. 
As one of the earliest examples of a RES incentive program, the German government introduced
the Renewable Energy Sources Act (EEG) in 2000. Among others things,
the act regulates the remuneration of RES, which are granted a technology-specific
compensation for each kWh of electricity fed into the grid. For photovoltaics,
the instrument has been effective in creating a dynamic demand and
a competitive supplier and installation industry \citep{Seel2014}. 

This feed-in tariff remuneration scheme is a remarkable possibility to study the
impacts of incentives on the observed deployment dynamics: The basic logic of the incentive program \textendash{} a fixed compensation
for 20 years starting with the date of initial operation \textendash{}
did not change for residential PV; the level of remuneration and system
costs, however, have changed. This allows to examine the effect of
this particular policy instrument by assessing the relationship between
profitability of PV systems and the aggregated deployment. 

\begin{figure}[H]
\includegraphics[width=12cm]{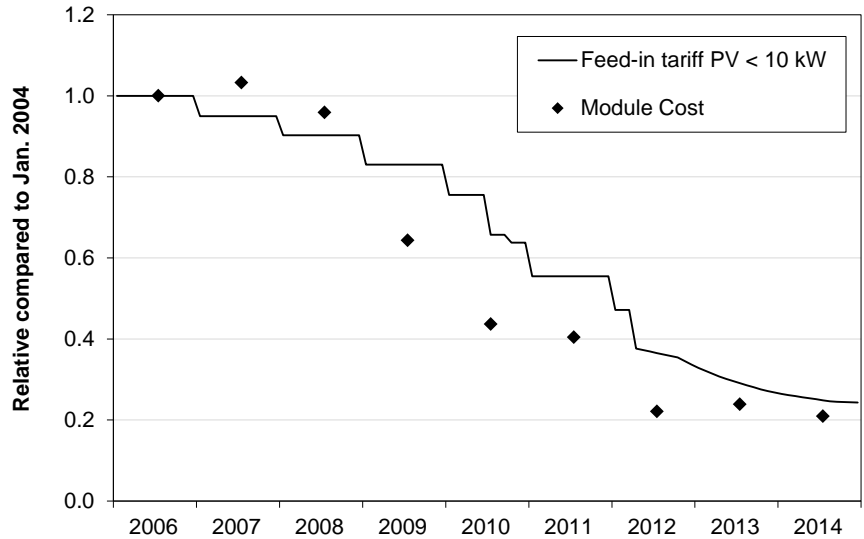}

\caption{\label{fig:Rel-dev-fit}\emph{Relative development of feed-in tariff
and PV module costs. }The feed-in tariff for residential PV systems
with a capacity of <10 $\textrm{kW}_{\textrm{p}}$ is shown by the
solid line. The markers indicate the PV module costs in the same period
of time. Both feed-in tariff and module cost developments are given
relative to their values in January 2006. The sharp decrease in PV
module cost made tariff adjustments necessary, and the developments
were not in alignment at all times. Data source: \citep{Bundesnetzagentur2016,Farmer2016}}
\end{figure}

Remuneration adjustments were necessary because the economics of PV
have been shifting rapidly \citep{Candelise2013}: PV module cost
decreased by approximately 80\% in the last 10 years alone \citep{Farmer2016}.
Figure \ref{fig:Rel-dev-fit} depicts the relative development of
PV module cost and feed-in tariffs for solar photovoltaic systems.
These developments were not in alignment at all times, especially
in the year 2009-2012. As module prices fell, remunerations were decreased,
often hastily, between 2006-2010 stepwise in a yearly way, between
2010-2012 in higher iterations as the rapid price decline made more
amendments necessary, and since April 2012 on a monthly basis in dependence
of the actual deployment over the past year. 

\begin{figure}[H]
\includegraphics[clip,width=12cm]{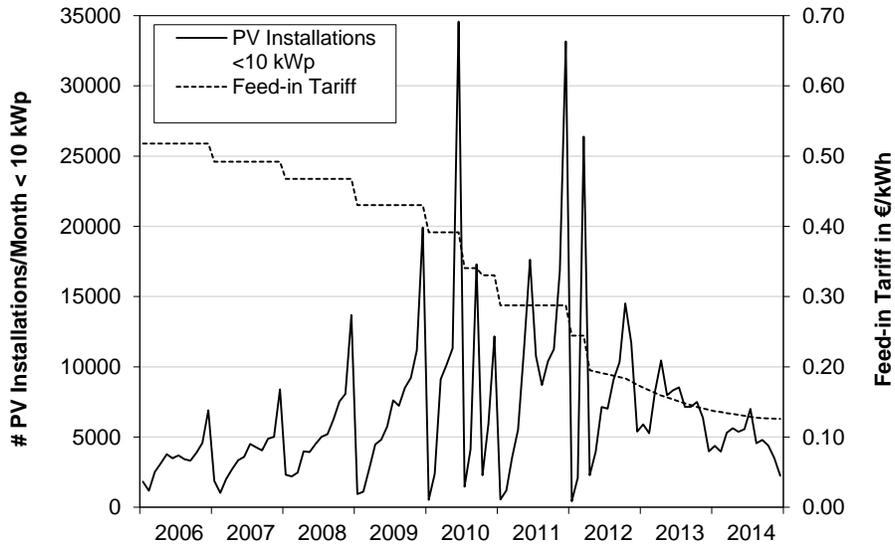}

\caption{\label{fig:deployment-FITs}\emph{Monthly PV installations and feed-in
tariff. }For the years 2006 to 2014, the installations of PV systems
<10 $\textrm{kW}_{\textrm{p}}$ per month are depicted by the solid
line, corresponding numbers are given on the left axis. The dashed
line shows values of the feed-in tariff given in Euro/kWh, scale on
the right axis. The installation peaks correspond with anticipated
feed-in tariff cuts. Data source: \citep{Bundesnetzagentur2016,OpenPowerSystemData2017}}
\end{figure}

Figure \ref{fig:deployment-FITs} illustrates the monthly PV installations
<10 $\textrm{kW}_{\textrm{p}}$%
\footnote{$\textrm{kW}_{\textrm{p}}$ is an often employed unit to depict the
nominal power of PV systems. It measures the output of a system
under peak (hence the \textquotedblleft{}p\textquotedblright{}) conditions,
i.e. standard testing conditions with a horizontal irradiance of 1
kW/m\texttwosuperior{} at 25\textdegree{}C ambient temperature.%
} between 2006 and 2014, in total about
700,000 installations. The development is characterized by pronounced
spikes, which correspond with anticipated step-wise feed-in tariff
cuts \citep{Leepa2013}.

\section{\label{sec:Methods}Methodology}

In order to reduce complexity, the study abstains from looking on individual
level decision making and focuses on the aggregate of investment dynamics.
To establish a link between the profitability and deployment, home-owners
are regarded to consider the installation of a PV system as an investment.
As such, PV systems have to compete with other investment possibilities.
With decreasing economy wide average rates of return, a lower internal
rate of return (IRR) on PV installations becomes more acceptable for
profit-oriented installers, as comparative investments on other markets
get less attractive. The modeling steps are outlined below and substantiated
in the following subsections:
\begin{enumerate}
\item Regard residential PV installations as investments, and calculate
retrospective net present values (NPVs) per time.
\item Derive a mean IRR in order to compare it to public-sector bond rates.
\item Compile a utility function $u(t)$ that increases exponentially with
increasing risk-adjusted IRR. Correlate this utility measure against
the deployment curve to see how well the model performs.
\item As an extension, consider changes in the utility function via the
value function of Prospect Theory.
\end{enumerate}

\subsection{Profitability modeling via NPV calculations}

Concerning investment choices, the net present value (NPV) method is often
used to decide whether to accept ($\textrm{NPV}>0$) and reject ($\textrm{NPV}<0$)
a project \citep{Brealey2000}. This method is used to assess the profitability
of PV systems over time. The NPV is calculated as follows: 

\begin{equation}
NPV\left(r,t\right)=-C_{0}\left(t\right)+\overset{T}{\underset{n=1}{\sum}}\frac{C_{+,n}\left(t\right)-C_{-,n}\left(t\right)}{\left(1+r\right)^{n}}\label{eq:1}
\end{equation}
where $C_{0}$, $C_{+,n}$, $C_{-,n}$ denote initial investment and
positive and negative cash flows, respectively, in the $n^{\textrm{th}}$ year after deployment
at time $t$. $r$ is the discount rate, $T$ the project lifetime. In
this case, the initial investment is given by

\begin{equation}
C_{0}\left(t\right)=s\cdot I\left(s,t\right)\label{eq:2}
\end{equation}

where $s$ denotes the system size in $\textrm{kW}_{\textrm{p}}$ 
and $I$ the specific investment cost per $\textrm{kW}_{\textrm{p}}$.
To account for higher specific installation cost for smaller installations,
initial investment cost is scaled to the system size $s$ according
to:

\begin{equation}
I\left(s,t\right)=I_{0}\left(t\right)\text{\ensuremath{\left(\frac{s}{10\,\textrm{kW}_{p}}\right)}}^{-0.063}\label{eq:3}
\end{equation}
The scaling parameter is derived from \citet{Feldman2012}, who provide
installation cost data by system size for installations in the year
2011 in the United States. $I_{0}\left(t\right)$ is the specific
investment cost for installations with a size of $10\,\textrm{kW}_{\text{p}}$.
Positive cash flows $C_{+,n}$ stem from feed-in tariff revenues and
avoided cost for grid electricity if part of the generated electricity
is self-consumed: 

\begin{equation}
C_{+,n}\left(t\right)=\begin{cases}
E_{n}\cdot f\left(t\right) & \textrm{if}\textrm{\,\ }f\left(t\right)>e\left(t\right)+f_{SC}\left(t\right)\\
E_{n}\left(f\left(t\right)\cdot\left(1-SC\right)+\left(e\left(t\right)+f_{SC}\left(t\right)\right)\cdot SC\right) & \textrm{else}
\end{cases}\label{eq:4}
\end{equation}
$E_{n}$ is the energy output of the system in year $n$, $f\left(t\right)$
the feed-in tariff at installation time $t$ and $e(t)$ the retail
electricity price. Between January 2009 and March 2012, roof-top PV
systems could receive an additional feed-in tariff for self-consumed
electricity $f_{SC}$$\left(t\right)$, which makes the case differentiation
in formula \ref{eq:4} necessary. $SC$ is the self-consumption ratio%
\footnote{Defined in \citet{Luthander2016} as ``(...) the share of self-consumed
electricity relative to total PV electricity production.''. Note
that the second case in formula \ref{eq:4} is also true if there
is no feed-in tariff for self-consumed electricity, but the standard
feed-in tariff is lower than the retail electricity price.%
}. The energy output $E_{n}$ of the system in year $n$ is calculated
as

\begin{equation}
E_{n}=s\cdot\gamma\cdot PR\cdot H_{opt}\cdot\left(1-d\right)^{n}\label{eq:5}
\end{equation}
$PR$ denotes the performance ratio of the system, $H_{opt}$ the
irradiance of an optimally inclined surface per m\texttwosuperior{}
and year in kWh%
\footnote{Since the rated capacity in $\textrm{kW}_{\textrm{p}}$ is defined
as the output under standard testing conditions (1 kW/m\texttwosuperior{}),
the irradiance in $\frac{\textrm{kWh}}{\textrm{m\text{\texttwosuperior}a}}$
can also be expressed in $\frac{\textrm{kWh}}{\textrm{kW}_{\textrm{p}}\textrm{a}}$,
so the units add up.%
}, and $d$ the degradation rate. The factor $\gamma$ can take values
from 0 to 1 and describes the roof\textquoteright{}s deviation from
the optimal inclination (with 1 being optimally inclined). For negative
cash flows $C_{-,n}$, only operation and maintenance costs are considered
and approximated with a yearly fixed share $c_{O\&M}$ of the initial
investment:

\begin{equation}
C_{-,n}=C_{0}(t)\cdot c_{O\&M}\label{eq:6}
\end{equation}

\subsection{Derivation of mean IRR - Monte Carlo Simulation }

Via a Monte Carlo method, the input parameters for the NPV calculation
are systematically varied. The method is described in detail by \citet{Darling2011}
for PV applications. In difference to most other NPV Monte Carlo simulations,
which are used for sensitivity analysis and risk assessment \citep{Hacura2001},
the study uses a slightly different interpretation by generating a set of
\emph{possible} systems which could be implemented in reality. With
each Monte Carlo iteration, a single NPV as described in formulas
\ref{eq:1}--\ref{eq:6} with a randomly drawn set of input parameters
is calculated. 

Given this input, an economic potential with respect to the discount
rate is derived. Here, the economic potential $\Theta$ is defined as the
share of acceptable possible projects (NPV > 0) 
as a function of the discount rate $r$ and deployment time $t$: 

\begin{equation}
\Theta\left(r,t\right)=N^{-1}\overset{N}{\underset{i=1}{\sum}}\delta_{i}\left(r,t\right),\;\delta_{i}(r,t)=\begin{cases}
1 & \textrm{if}\textrm{\,\ NPV}>0\\
0 & \textrm{if}\textrm{\,\ NPV}\leq0
\end{cases}\label{eq:7}
\end{equation}
where index $i$ denotes a single Monte-Carlo sample and $N$ the
total amount of samples (100,000 in our study).

By calculating $\Theta\left(r,t\right)$, i.e. the positive share
of NPVs, for different discount rates $r$ (in this study from -10.0\%
to +15.0\% with a step size of 0.5\%), the mean internal rates of
return (IRR) can be identified: The IRR is defined by the rate $r$
for which the NPV is exactly 0. As can be seen in formula \ref{eq:1},
NPV decreases with increasing $r$ (the denominator of the function
gets larger) if the cash-flows in the sum are positive, which is the
case as the operation and maintenance costs $C_{-,n}\left(t\right)$
are relatively small. Hence $\Theta\left(r,t\right)$ decreases monotonically
if the discount rate $r$ is increased. Some particular possible installations
will change sign of NPV from positive to negative with each increasing
discount rate step. The retrieved rate $r$ is the particular IRR
for those installations.

\begin{figure}[H]
\includegraphics[bb=0cm 5cm 595bp 420bp,width=12cm]{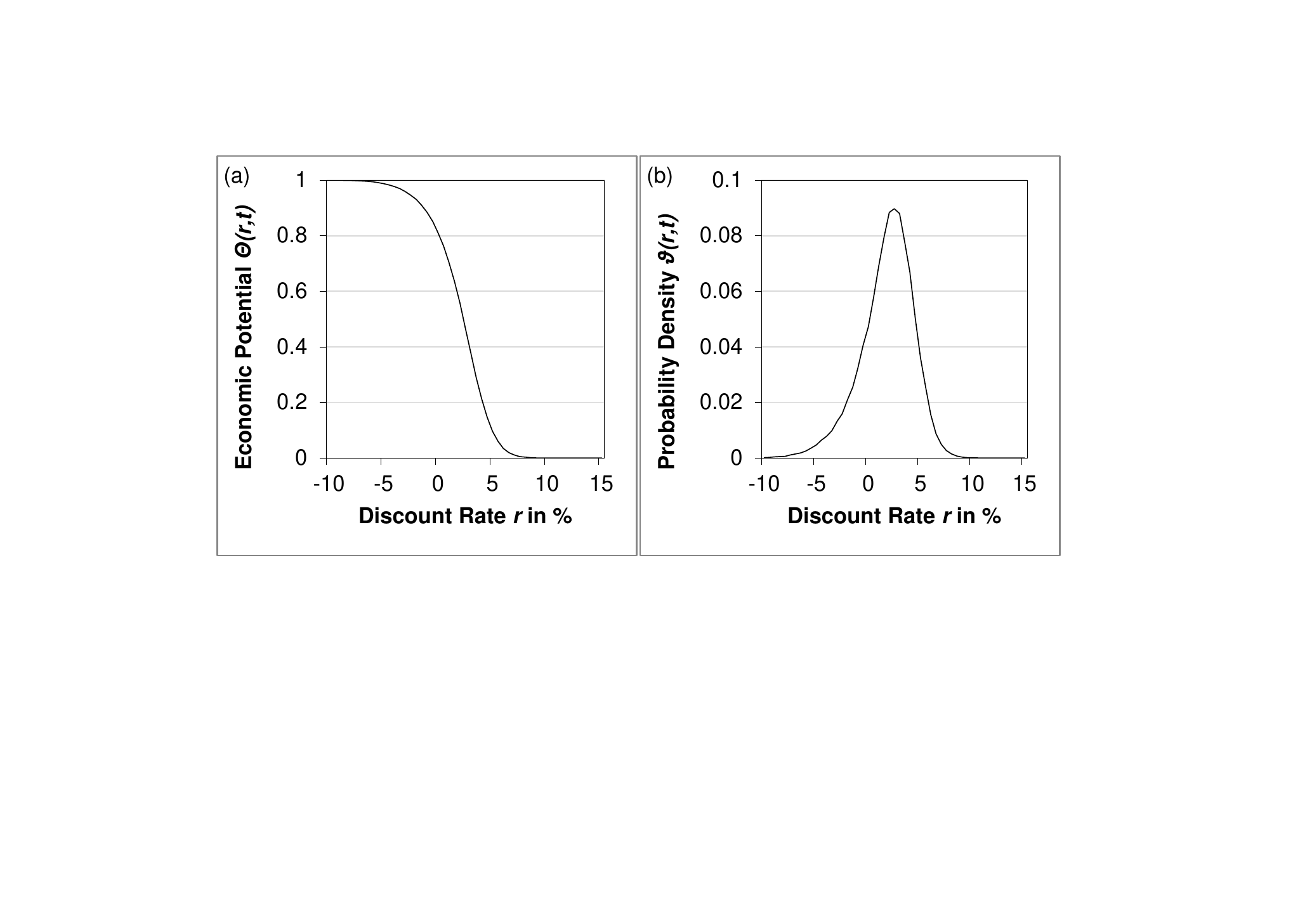}

\caption{\label{fig:Example-econ-pot}\emph{Example for the Economic Potential
$\Theta\left(r,t\right)$}, i.e. the share of positive NPVs, and its
derivative, the Probability Density Function of IRRs $\vartheta\left(r,t\right)$
for t = Dec. 2008.}
\end{figure}

Hence, the derivative of $\Theta\left(r,t\right)$ with respect to
$r$, $\vartheta\left(r,t\right)=-\tfrac{\partial\Theta\left(r,t\right)}{\partial r}$,
yields the probability density function of IRRs for all projects considered
in the simulation, see figure \ref{fig:Example-econ-pot} for a graphical
representation. It can be interpreted by the slope of the economic
potential \citep{Hillier1963}. The mean IRR value can then be easily
extracted from the density function $\vartheta\left(r,t\right)$:

\begin{equation}
\overline{IRR}\left(t\right)=\int r\cdot\vartheta\left(r,t\right)dr\label{eq:8}
\end{equation}
In this study, the continuous case was not considered but approximated
with a stepwise integration for finite increments of $\Delta r$:

\begin{equation}
\overline{IRR}\left(t\right)=\underset{r}{\sum}r\left(\Theta\left(r,t\right)-\Theta\left(r+\Delta r,t\right)\right)\label{eq:9}
\end{equation}

The derivation of an economic potential via a Monte Carlo simulation
of a broad set of possible systems ensures that the economic assessment
is as unbiased as possible. Calculating only a single reference system
with fixed parameters like irradiation and size could skew the profitability
analysis, since the inputs can potentially have non-linear effects
on profitability over time. The relative economic prospect of a single
PV system could shift from more to less favorably or vice versa (it
could be, for instance, that small PV systems are relatively better
off in earlier moments of the analysis). Calculating a mean IRR in
the way presented can lessen this problem, as a broad share of possible
PV systems is considered.

\subsection{Deployment modeling via a utility function}

The objective of the following two sub-sections is to find some utility
measure that correlates well with the observed deployment of residential
PV systems. 

\begin{figure}[H]
\includegraphics[width=12cm]{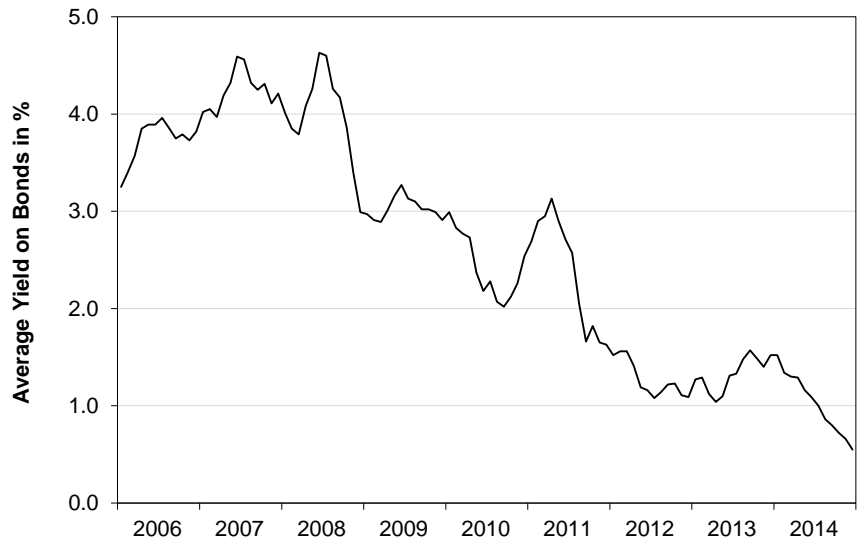}

\caption{\label{fig:Risk-free-rate}\emph{Risk-free rate of return in Germany
over time}, which is derived from the average yield of public-sector
bonds. Data source: \citet{DeutscheBundesbank2016}}
\end{figure}

PV system are regarded as an investment that has to compete with economy
wide average rates of return. The rate of public-sector bonds has
changed considerably: The average return on German government bonds,
which is considered to be the risk-free alternative investment
for the purpose of our evaluation (abbreviated as $\rho\left(t\right)$),
have been decreasing from nearly 4-5\% in the year 2006-2008 to less
than 1\% in the year 2014 in the aftermath of the financial crisis
of 2007-08 and the turmoil on European markets. Figure \ref{fig:Risk-free-rate}
shows the average yield of public-sector bonds in Germany over time. 

The risk-adjusted IRR $\pi\left(t\right)$ shall be defined as

\begin{equation}
\pi\left(t\right)=\overline{IRR}\left(t\right)-\rho\left(t\right)\label{eq:11-1}
\end{equation}

In the case of continuous interest compounding, the payout $P$ after
time $t$ subject to a return rate $r$ can be expressed as an exponential
function \citep{Brealey2000}: 

\begin{equation}
P\varpropto e^{rt}\label{eq:12-5}
\end{equation}

Hence, and similar
to the fit functions used by \citet{Benthem2008} and \citet{Wand2011},
it is assumed that consumers get an exponentially increasing utility
$u\left(t\right)$ with an increasing rate of the risk-adjusted 
IRR $\pi\left(t\right)$ (the return rate $r$ in formula \ref{eq:12-5}) if they choose to invest at time $t$:

\begin{equation}
u\left(t\right)=e^{\kappa\pi\left(t\right) }\label{eq:12-2}
\end{equation}
As the feed-in remuneration is paid out for 20 years in Germany, this time-frame (i.e. the economic lifetime $T$) is taken as value
for $\kappa$. The deployment $d(t)$ is modeled to be proportional
to the aforementioned exponential utility function $u\left(t\right)$,
with a constant $c$ to be determined by the actual uptake:

\begin{equation}
d\left(t\right)=c\cdot u\left(t\right)\label{eq:13-1}
\end{equation}

\subsection{Prospect Utility Model}

In a seminal paper, Kahneman and Tversky introduced \emph{Prospect
Theory} to the scientific community \citep{Kahneman1979}. Central
to the theory is the idea that people often do not perceive utilities
in absolute values of wealth, but rather in gains and losses relative
to the current state. Moreover, as they phrase it, \textquotedblleft{}losses
loom larger than gains\textquotedblright{} \citep{Kahneman1979},
which means that the disutility of a loss is perceived worse than
the utility of a gain of the same absolute size. Prospect theory has
become an integral part of behavioral economics, and has been successfully
applied to problems in finance and insurance, among others \citep{Barberis2013}. 

\begin{figure}[H]
\includegraphics[width=12cm]{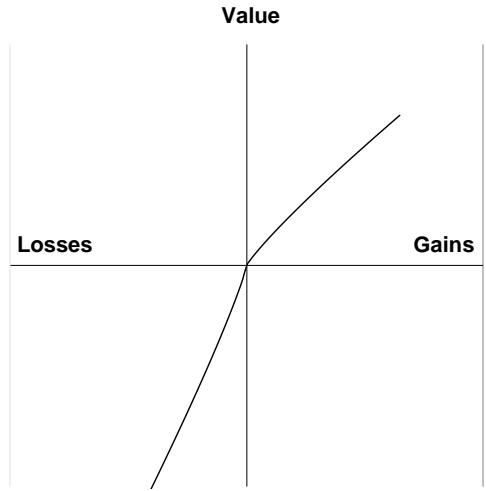}

\caption{\label{fig:The-value-function-1}\emph{The value function of prospect
theory.} The disutility of losses is comparatively larger than the
utility of gains of the same absolute size. The shape of the value
function can be measured experimentally. For example, given a gambling
game (like the toss of a coin) with 50\% chance of losing \$100, most
people start to accept the game if they have a 50\% change of winning
at least \$200 or more \citep{Tversky1992} -- a phenomenon far too large to be explainable
by income effects \citep{Tversky1992}. }
\end{figure}

The theory consists of two parts; a value function which assigns values
to gains and losses relative to the status quo, and a weighting function
which is used to assign weights on how people perceive probabilities
(people have consistently been shown to misjudge very small and very
large probabilities). The value function (for a graphical depiction
see figure \ref{fig:The-value-function-1}) has a kink at the origin,
meaning that relative losses cause a higher disutility (seen in the
larger slope for losses compared to gains), an effect they coin with
the phrase \textquotedblleft{}loss aversion\textquotedblright{} \citep{Tversky1992}.
Moreover, there is some \textquotedblleft{}saturation\textquotedblright{}
in the value curve -- the slope decreases for values further away
from the origin, i.e. people are more indifferent about a marginal
win or loss far away from the status quo than near to it. 

The theory is applied to the PV investment problem in the following
way: It is postulated that potential residential PV adopters do not only
rate the investment's attractiveness in absolute terms (i.e. in terms
of utility derived from risk-adjusted IRR), but also in relative gains
and loss, i.e. in the frame of \emph{changes} of that utility function.
Imagine policy makers announce to lower remunerations (the legislative
plans are usually revealed several months in advance). Potential adopters
realize a prospective PV system would have less profitability than
as of today. In the light of loss aversion, this would be a further
incentive to build in order to avoid the disutility of this potential
loss, even if the absolute profitability is comparatively average. 

The value function of prospect theory is parametrized as follows \citep{Tversky1992}:

\begin{equation}
v\left(x\right)=\begin{cases}
x^{\alpha} & \textrm{if}\, x>0\\
-\lambda\left(-x\right)^{\alpha} & \textrm{if}\, x\leq0
\end{cases}\label{eq:10}
\end{equation}
where $x$ is a relative gain ($x>0$) or loss ($x<0$), $\alpha$
is the saturation parameter and $\lambda$ the extent of the loss
aversion in comparison to gains. They experimentally found $\alpha=0.88$
(i.e. only minor saturation) and $\lambda=2.25$ (i.e. losses are
perceived 2.25 times as badly as gains of the same absolute extent).

This value function is used directly to model the investment dynamics.
One major difficulty in applying the value function is to create a scale
for gains and losses \citep{Barberis2013}. In our example, gains
and losses are defined in terms of prospective changes in the exponential
utility function $u\left(t\right)$%
\footnote{A time step size $\Delta t=1\,\textrm{month}$ is used, as data resolution
and feed-in tariff adjustments since 2012 have the same step size.%
}: 
\begin{equation}
x_{\rightarrow}\left(t\right)=u\left(t+\Delta t\right)-u\left(t\right)=e^{\kappa\pi\left(t+\Delta t\right)}-e^{\kappa\pi\left(t\right)}\label{eq:11}
\end{equation}
Additionally, one can also consider the retrospective changes in the
exponential utility function:

\begin{equation}
x_{\leftarrow}\left(t\right)=u\left(t\right)-u\left(t-\Delta t\right)=e^{\kappa\pi\left(t\right)}-e^{\kappa\pi\left(t-\Delta t\right)}\label{eq:11-2}
\end{equation}
A prospect utility function $U\left(t\right)$ of investing
in PV is proposed, which comprises the exponential utility function $u(t)$, minus
the utility of its forward change (an expected loss in the next month
will increase deployment)%
\footnote{Note that a prospective \emph{dis}utility of not investing
in PV is interpreted as an incentive to build, therefore the minus sign in formula
\ref{eq:11}.%
}, plus the utility of its backward change (a loss compared to the
last month will decrease deployment):
\begin{equation}
U\left(t\right)=u\left(t\right)-v\left(x_{\rightarrow}\left(t\right)\right)+v\left(x_{\leftarrow}\left(t\right)\right)\label{eq:12}
\end{equation}

with $x\left(t\right)$ from formula \ref{eq:11} and \ref{eq:11-2}
and the value function of prospect theory $v(x)$ from formula \ref{eq:10}. The parametrization of the value function is left unchanged from the original source \citep{Tversky1992}%
\footnote{For model simplicity, conditions under certainty are assumed in the present study.
Prospect theory was originally developed to assess decision under
uncertainty including so-called decision weights (either very small
or very large probabilities were shown to be poorly assessed by study
participants) \citep{Kahneman1979}. 
However, the original authors of the theory have shown that the value function
can also be applied in conditions under certainty in the same way \citep{Tversky1991}.%
}. 

The deployment per month is modeled to be proportional to the aforementioned
prospect utility function $U\left(t\right)$, with a constant $k$
to be determined by the actual uptake%
\footnote{It is assumed that the deployment would not go to negative values
if the prospect utility function went below 0.%
}:

\begin{equation}
d\left(t\right)=\begin{cases}
k\cdot U\left(t\right) & \textrm{if}\, U\left(t\right)>0\\
0 & \textrm{else}
\end{cases}\label{eq:13}
\end{equation}

\subsection{\label{sub:Data}Data }

Table \ref{tab:Input-parameters} presents an overview of the assumptions
for the distributions of the Monte Carlo calculation input parameters
in the study, and the respective data sources from where they are
derived. By model definition, all system sizes between >0 and $10\,\textrm{kW}_{\textrm{p}}$ are
equally probable, although the scaling function described in formula
\ref{eq:3} puts a price tag on smaller systems. For model simplicity,
a correlation between system size and the self-consumption ratio is
not considered. The irradiance distribution is derived from openly
accessible radiation maps \citep{Huld2012}. Roof inclination factors
are derived from \citet{Mainzer2014}. Both distributions are approximated
with beta functions, which can be fully characterized by its minimum,
maximum and modal value \citep{Davis2008}. 

For the installation cost $I_{0}$ between the last quarter of 2006
and 2014 a commercial dataset is used \citep{EuPDResearch2016}. The
data describes PV system costs (turnkey ready including modules, inverter
etc.) for systems $<10\,\textrm{kW}_{\textrm{p}}$ in 3-monthly resolution, which
was linearly interpolated to get monthly values. PV deployment data
is taken from sources of the Open Power System Data project \citep{OpenPowerSystemData2017},
which builds on sources published by the transmission system operators
and the network regulator \citep{50HertzTransmission2016,Amprion2016,Tennet2016,TransnetBW2016,Bundesnetzagentur2016}.
It is assumed that all of the installations $<10\,\textrm{kW}_{\textrm{p}}$ 
were roof-top installations%
\footnote{This is a necessary assumption, as the installation data published by the transmission system operators is
not reliably differentiated between roof-top and ground-mounted PV systems.
However, ground-mounted PV systems tend to be much larger (near the
$\textrm{MW}_{\textrm{p}}$ range), so error should be minor, although
unknown in magnitude. %
}. The data contains entries for all RES installations which are incentivized
via the EEG, and includes date of installation, state, capacity, and
the respective distribution system operator, among others. The data
was filtered for PV installations $<10\,\textrm{kW}_{\textrm{p}}$ to get the
absolute number of monthly installations. Likely duplicates were removed
before processing. No other alterations were required. 

The risk-free rate is derived from the average yield on public-sector
bonds, data was obtained via \citet{DeutscheBundesbank2016}. Feed-in
tariff levels were obtained from \citet{Bundesnetzagentur2016}, for
an overview see \citet{SFV2016}.

Since socio-economic and environmental parameters like income and
irradiance are fairly homogeneously distributed over Germany and did
not change substantially over time, and data availability and the
number of installations are high, the assessment of the investment
dynamics is ensured to be as undistorted as possible. The analysis
is restricted to the years 2006-2014: Due to data availability, the
analysis starts in 2006. The years after 2014 are not considered because
of the start of the adoption of PV battery systems \citep{Kairies2015},
which change the economics of PV systems considerably \citep{Hoppmann2014}.
Changes in the number of available roofs were not considered, and
the break condition of investigated systems of $<10\,\textrm{kW}_{\textrm{p}}$ is rather arbitrary;
better data on both ends would reduce uncertainty regarding the actual
uptake of residential photovoltaic systems.

\section{\label{sec:Results}Results}

\begin{figure}[H]
\includegraphics[width=12cm]{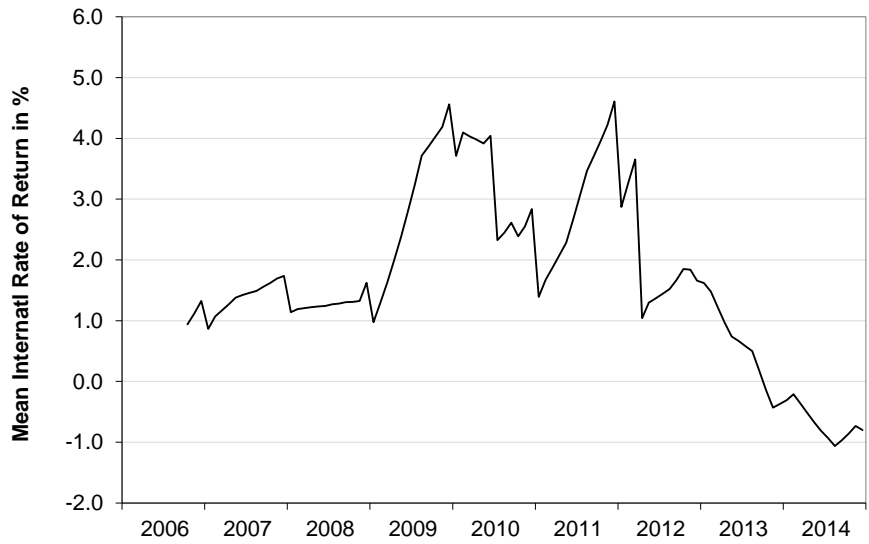}

\caption{\emph{\label{fig:Calculated-mean-IRR}Calculated mean IRR} of potential
residential photovoltaic systems in Germany. System cost and remuneration
are the fundamental determinants of profitability.}
\end{figure}

Figure \ref{fig:Calculated-mean-IRR} illustrates the calculated mean
IRR of possible residential photovoltaic systems over time. Since
system costs and remuneration are the fundamental determinants of
profitability, this graph relates to figure \ref{fig:Rel-dev-fit}:
The steps in the years 2006-2008 results from the yearly step-wise
adjustment of the feed-in tariff. The profitability of PV installations
rose significantly in the following years. Highest returns were possible
at the end of 2009 and 2011, respectively, right before remuneration
cuts. After further feed-in tariff adjustments in 2012, mean negative returns
where observable (note that a mean IRR of less than 0\% does not mean
that there is no incentive to build at all; there might be installations
well above 0\%, since this is only the mean value for all possible
systems). The IRR alone is only moderately correlated with deployment
(Pearson correlation of 0.47).

\begin{figure}[H]
\includegraphics[width=12cm]{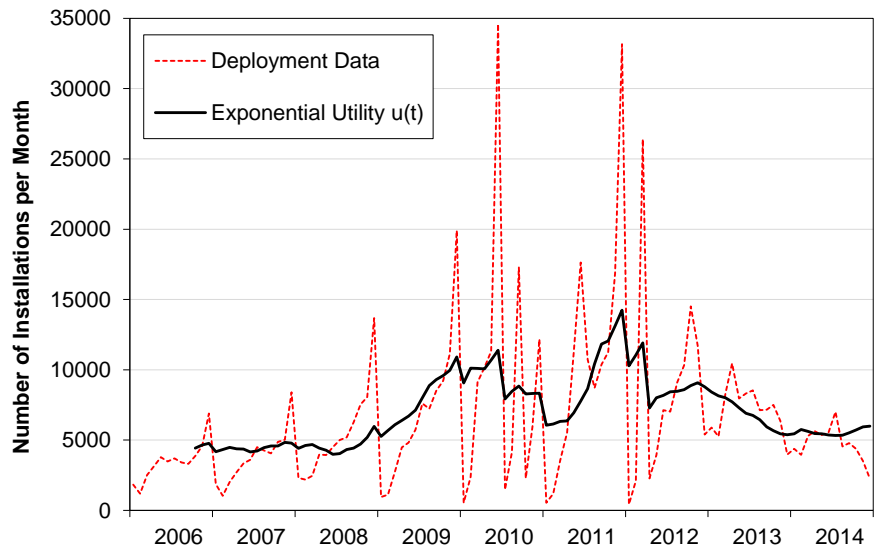}

\caption{\label{fig:Monthly-deployment-exp-fit}\emph{Data and exponential
utility model results for the monthly number of PV installations.}
The dashed line shows the observed number of monthly installations
of residential PV systems. It is superimposed with the results of
the fit of the exponential utility function to the data, indicated
by the solid line. The dynamics of the sub-yearly peaks of installations
are insufficiently represented. }
\end{figure}

Figure \ref{fig:Monthly-deployment-exp-fit} shows the fit of the
exponential utility model $u(t)$. The overall shape of the deployment
curve is covered, but the dynamics of the sub-yearly peaks of installations
are insufficiently represented. A moderate Pearson correlation of
0.62 is obtained ($p<0.001$). The scaling factor is empirically found
to be $c=7845.5$ to match with the absolute deployment over the time-frame
under consideration.

\begin{figure}[H]
\includegraphics[clip,width=12cm]{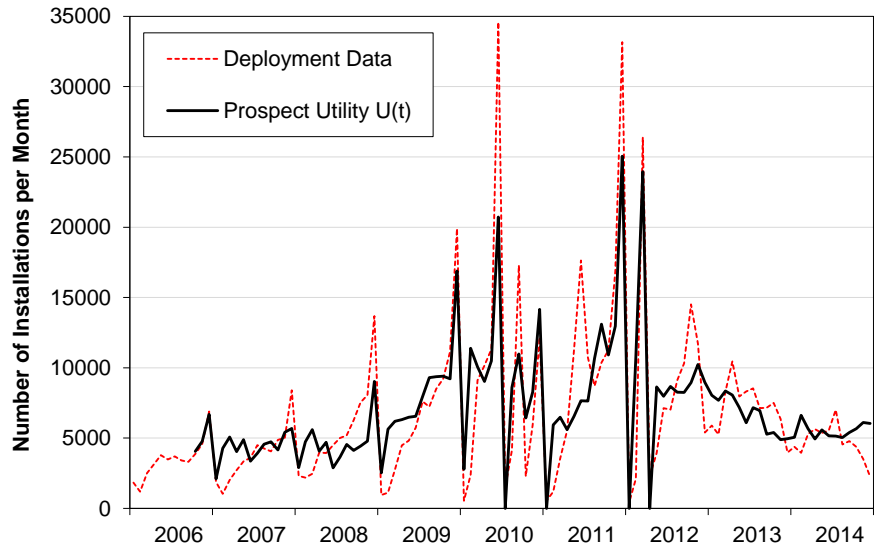}

\caption{\label{fig:Monthly-number-PV-combined-utility}\emph{Data and prospect
utility model results for the monthly number of PV installations.
}The dashed line shows the observed number of monthly installations
of residential PV systems. The superimposed solid line is the fit
of the prospect utility function to the data, i.e. the exponential
utility function combined with the value function of prospect theory.
The stylized features of the deployment curve are represented substantially
better.}
\end{figure}

The goodness of fit can be substantially improved if the value function
of prospect theory is incorporated into the evaluation (figure \ref{fig:Monthly-number-PV-combined-utility}).
The overall shape of the deployment curve with its stylized features is represented well, most notably the pronounced peaks and valleys. 
Not all peaks are met precisely in their height. Nevertheless, the location of the
peaks is almost always found. A high Pearson correlation of 0.85 is
obtained ($p<0.001$). One notable exception is the observed deployment
peak in mid 2011; its probable origin will be examined in the discussion
section. The scaling factor for this case is empirically found to
be $k=7516.2$. Table \ref{tab:Pearson-correlation-coefficients} gives an overview of the fitting results.

\section{\label{sec:Discussion}Discussion}

\subsection{Sensitivity Analysis}

\begin{figure}[H]
\includegraphics[clip,width=12cm]{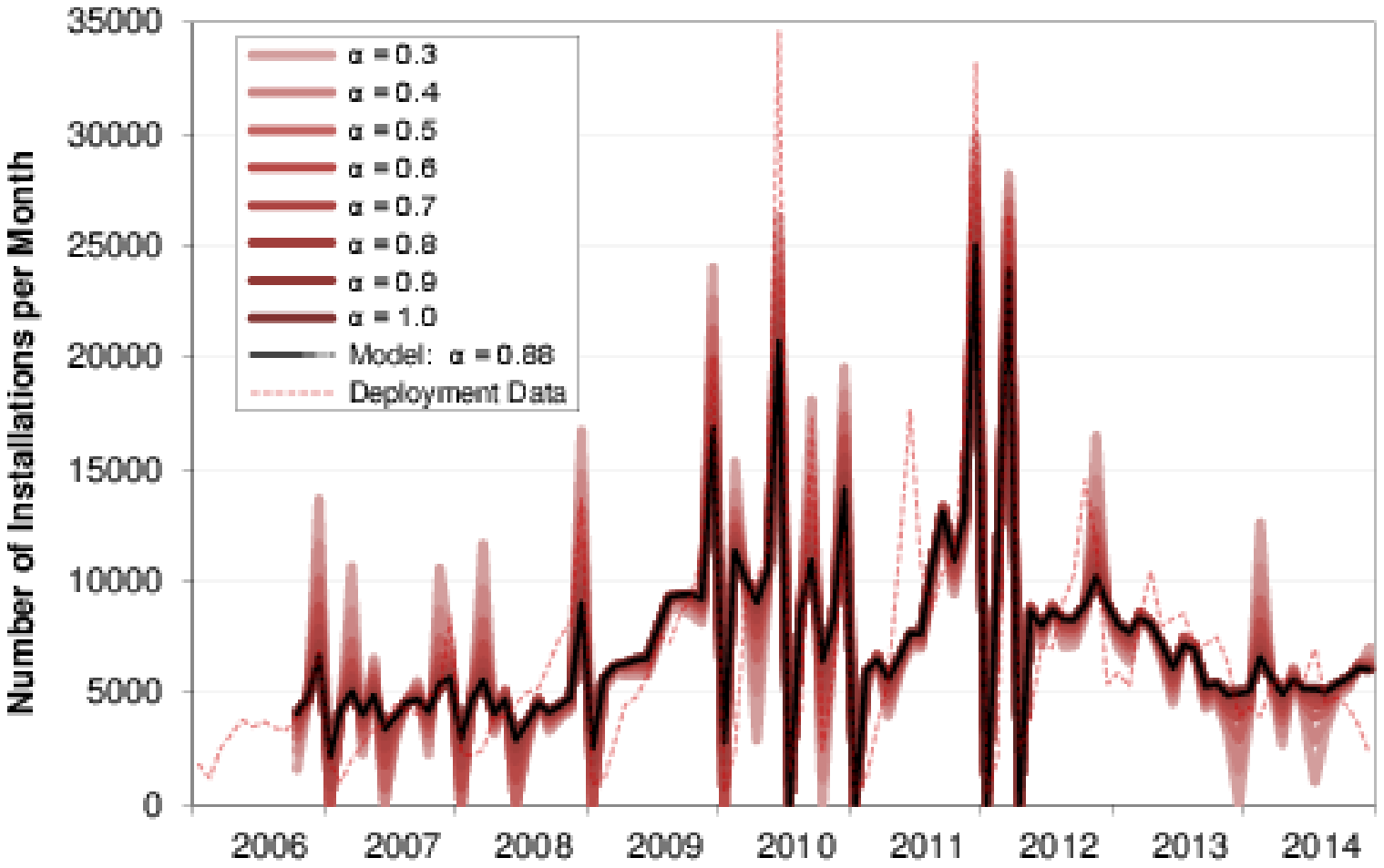}

\caption{\label{fig:sensitivity_alpha}\emph{Sensitivity of the numerical value of $\alpha$ from the value function of prospect theory (formula \ref{eq:10}) on the uptake model.
}}
\end{figure}

\begin{figure}[H]
\includegraphics[clip,width=12cm]{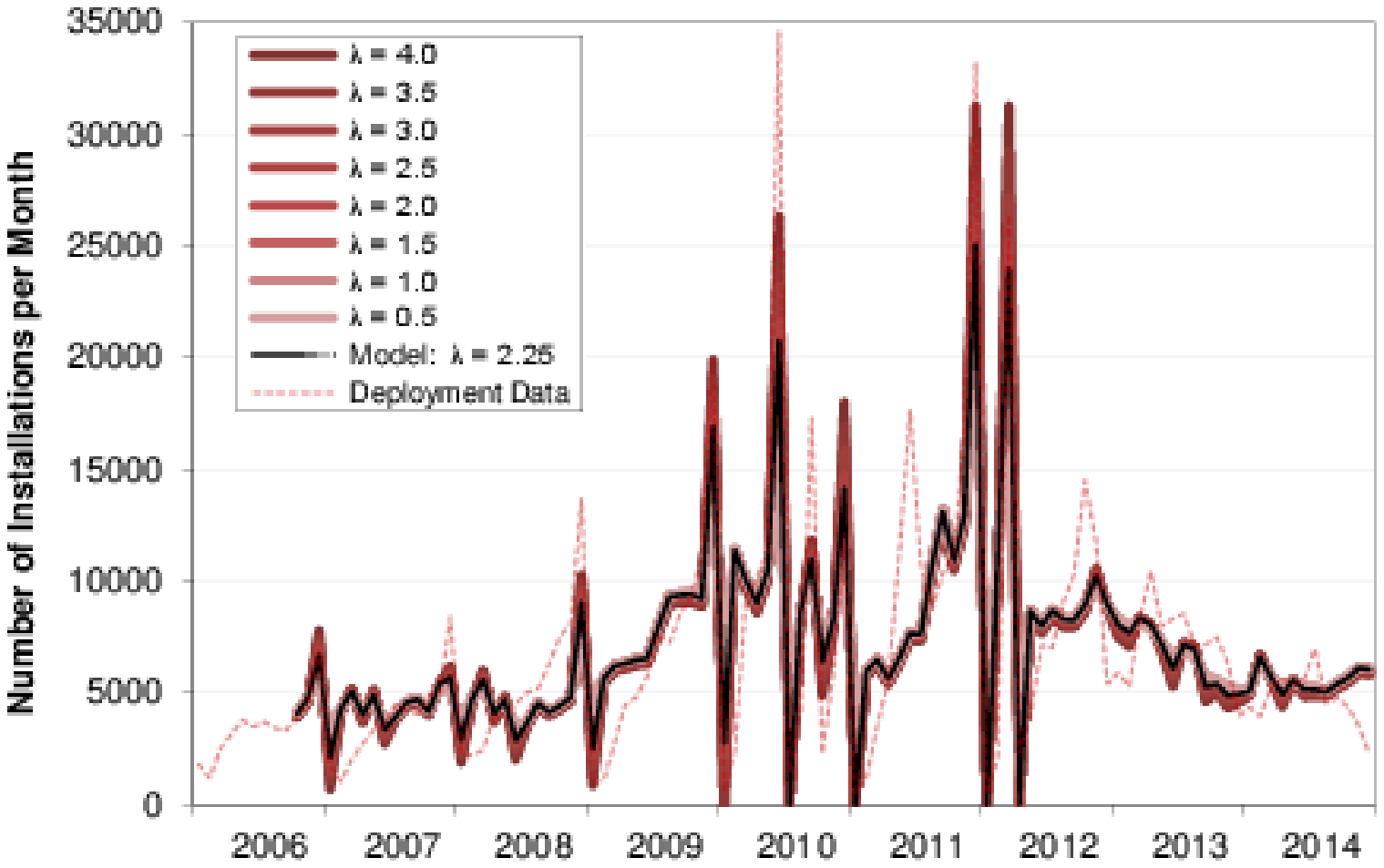}

\caption{\label{fig:sensitivity_lambda}\emph{Sensitivity of the numerical value of $\lambda$ from the value function of prospect theory (formula \ref{eq:10}) on the uptake model.
}}
\end{figure}

Note that the results were not subject to careful parameter adjustment.
Fitting is only performed in the last step of the analysis to translate
the relative utility scales to absolute deployment levels. This is
remarkable because the value function has been parametrized in a completely
different context (lab controlled gambling games \citep{Tversky1992})
and is applied with unchanged parametrization to this case. 
To study the influence of this implicit parameter choice, a sensitivity analysis of the functional parameters of the value function ($\alpha$ and $\lambda$, see formula \ref{eq:10}) is performed. The sensitivity plots are shown in figures \ref{fig:sensitivity_alpha}-\ref{fig:sensitivity_lambda}.

Lower numerical values for $\alpha$ correspond to a higher saturation of the value function -- the magnitude of the value change becomes less relevant.
Hence, smaller fluctuations in the exponential utility function $u\left(t\right)$ lead to comparatively higher deployment peaks and valleys (see for example the years 2006-2008, figure \ref{fig:sensitivity_alpha}).
For very small values of $\alpha < 0.5$, the model represents the deployment data poorly.

The extent of loss aversion is parametrized with $\lambda$. If, for example, losses are perceived twice as badly as gains of the same absolute extent, $\lambda$ would take the value of 2. With higher levels of $\lambda$, the deployment peaks and valleys become more prominent in the uptake model (figure \ref{fig:sensitivity_lambda}), as losses would be perceived comparatively higher. Overall, the model is rather robust towards changes in the parameter $\lambda$, the stylized features of the deployment curve are represented well for a broad window of parameters.

\begin{figure}[H]
\includegraphics[clip,width=12cm]{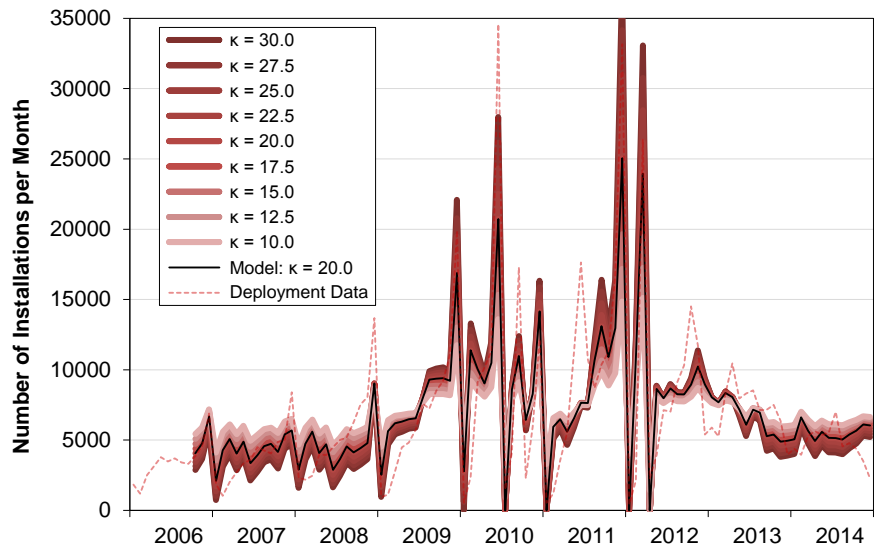}

\caption{\label{fig:sensitivity_kappa}\emph{Sensitivity of the numerical value of $\kappa$ from the utility function (formula \ref{eq:12-2}) on the uptake model.
}}
\end{figure}

Additionally, a sensitivity analysis of parameter $\kappa$ of the exponential utility function (see formula \ref{eq:12-2}) is performed. The plot is shown in figure \ref{fig:sensitivity_kappa}.
Higher values of  $\kappa$ correspond to comparatively higher utility values for higher risk-adjusted returns $\pi (t)$. Hence, with higher levels of $\kappa$, comparatively more uptake is predicted when risk-adjusted returns are higher (mostly between 2009-2012).

\subsection{Interpretation of the results}

The data shows that investments increase right before a reduction
of the remuneration - a policy change induces a strengthened uptake.
This notion can be extracted out of the deployment curve without further
numerical analysis (see figure \ref{fig:deployment-FITs}). Peak deployment
dynamics are represented via the frame of gains and losses of investors
based on Kahneman and Tversky's value function. Together with a common
net present value calculation, this seems to fill the explanation
gap of the observed German PV installation data. 

The sensitivity analysis reveals that the model is rather robust towards the parametrization 
of the value function and the exponential utility function.
Therefore, the presented approach is considered ``fundamental'' in a sense that
it can be directly derived from fundamental economic considerations (like the utility
from compound interest over 20 years) or from behavioral experiments 
(like shape of the value function). 

It is probable that most people do not account and reason about PV
in the way as it has been presented in the paper (it has in fact been
shown by \citet{Salm2016} and several other studies that a large
share of PV adopters rely on ``gut feelings\textquotedblright{} and
simple heuristics like payback times). Is it therefore reasonable
to focus solely on the described (behavioral) economic aspects? The
proposed theoretical approach does not cover possible word of mouth
effects or even personal values of investors. Though internal factors
of decision makers play a role in general and may also interact and
influence with external factors as \citet{Kastner2016} suggest, the
study evaluates financial incentives solely. Thus, the dataset and
methodology does not allow for conclusions about the effect of internal
factors (like personal norms, values and attitudes) on the investment
decision. However, it is likely that improved economics will convince
more people to adopt, no matter if they are convinced environmentalists
or if they view photovoltaics just as an attractive investment among
others, as the financial attractiveness should shift the attitude
concerning investment of all actor groups in the same direction. The
study can therefore help to understand how the economic prospect and
its changes affect the overall adoption pattern. In a sense, it is
remarkable how \emph{much} one can explain with economic prospects
alone.

\subsection{\label{sub:Future-work}Possible extensions of the study}

The presented method can be used for forecasting. The prospect utility
 model might help policy makers find appropriate remuneration levels
to reach desired deployment goals. For example, the study could be
used to craft exploratory energy scenarios which depict the uptake
of future energy technology combinations like PV battery systems.
Additionally, the study helps to understand the influence of bond
rates on deployment levels -- in agreement with \citet{Leepa2013},
it could be shown that the economic assessment largely benefits if
the risk-free return rates are factored in. For example, PV deployment
would have been lower according to the model if interest rates had
remained at above financial crisis levels. This paper, however,
abstains from using the model in a forecasting or scenario fashion
as the results should get validated with a different case study first. 

Future work should examine and apply this method to other cases and
countries or to other kinds of (institutional) investors to see if
those data patterns are still observable. The proposed model should
be indifferent to remuneration policy instruments, i.e. should not be limited to
feed-in tariff schemes, as the NPV calculation method is by definition
only subject to cash-flows, irrespective where they come from (purchasing
agreements, tax credits, etc.). Furthermore, the method should be
applicable to any kind of energy investments, if the investment in
question is not mutually exclusive but to the standard risk-less choice
of bonds, and has short installation times and low running cost. This
means on the contrary that other energy technologies with long project
development times like wind power will be harder to depict as the
time gap between investment decision and implementation will be higher.

Furthermore, one could investigate socio-economic or spatial aspects
of the described effect. Possible investors might become aware of
remuneration schemes via word of mouth effects or media representation
of the topic. The authors cannot estimate the influence of peer to
peer effects with the given data; investigating the impact of these
effects on the investment dynamics would require a different research
design in order to compellingly work out possible correlation patterns.
However, some aggregated information about the overall ``likelihood''
of investment irrespective of financial concerns is contained in the
scaling factor $k$, which is determined as a final step to link the
prospect utility measure and the absolute observed deployment (see
formula \ref{eq:13}). The value $k$ determines how much uptake is
obtained given ceteris paribus economic conditions; it can potentially
change over time if the attitude towards the technology changes. One
of the installation peaks can be an indication of this happening:
The installation peak in mid 2011 is not covered by the prospect utility
 model. This peak coincides with the Fukushima disaster and a public
attitude shift towards renewable energy in Germany. Future studies
could look into temporal and spatial aspects of the fitting value
$k$. 

There might be a fundamental explanation other than ``irrational''
loss aversion for the pronounced peak structure: Option value. The option 
value framework considers uncertainty as a main factor influencing investment
decisions. The larger the uncertainties, e.g. development of electricity prices or 
changing policy incentives, the less likely the investment. Thus households
would delay a favorable investment in order to ``buy time\textquotedblright{}
and to wait until uncertainty is resolved or better investment opportunities
arise \citep{Bauner2015}. 
In the context of the option value framework, an expected decrease of remuneration of
electricity from PV could lead to an increased uncertainty of future profits, 
as they depend more and more on volatile market prices. This might raise 
investments before a change of remuneration.
Vice versa, an expected increase of remuneration could delay investments.
So option value can also describe the effect of policy changes on the 
uptake of household PV investments.

The option value framework however considers rational decision making based on uncertainties 
and thus differs significantly from the proposed description based on the boundly rational perception
 of anticipated losses (which is substantially more myopic as only changes of utility one month in advance are considered). 
Further studies could look into the differences of both approaches in a 
more structured way to get deeper insights into their pros and cons and to which extent real world residential energy decisions
 are rationally grounded.

\section{\label{sec:Conclusion}Conclusions and Policy Implications}

The presented paper offers a new perspective on the residential PV
investment dynamics in Germany and successfully combines a NPV analysis
with prospect theory. Although the decision whether and when to invest
in photovoltaic systems is influenced by many factors, the selected
approach could reproduce most of the dynamics of the uptake with only
a few financial and behavioral assumptions. There are only few widely
accepted applications of prospect theory in economics \citep{Barberis2013}
-- the proposed model is one of the first numerical applications in
the energy sciences to the authors\textquoteright{} knowledge. The
proposed approach requires only one fitting parameter and is thus
fundamental and parsimonious enough to be incorporated into whole
system energy studies. 

The study is useful for policymakers in several ways. A better understanding
of the impact of deployment policies can help to design more robust
remuneration schemes. By and large, the observed deployment of residential
photovoltaics in Germany can be explained by the anticipation of profitability,
and most notably, additionally by its anticipated change. 

Stepwise changes in the remuneration design can therefore induce non-linear and non-intended
investment behavior. According to the model, this effect is temporary however, and 
poses only a problem if a narrow window of uptake is considered when re-adjusting the height of remunerations over time.

\pagebreak{}

\bibliographystyle{elsarticle-harv}
\bibliography{arxiv}

\pagebreak{}

\section*{Acknowledgments}

We would like to thank Ulrich Frey, Kristina Nienhaus, Matthias Reeg,
André Thess and Laurens de Vries for fruitful discussions about this
work. We also benefited from comments by members of the Helmholtz
Research School on Energy Scenarios, the ETH Zürich PhD Academy on
Sustainability and Technology in Appenzell (Switzerland) in 2015,
the 14th Symposium Energy Innovation in Graz (Austria) and the 5th
BAEE Research Workshop on Energy Economics in Delft (The Netherlands)
in 2016. Dominique Heiken provided valuable research assistance. 
We thank François Lafond for providing data on PV module costs.
The study was financed by the basic funding of DLR, which we kindly acknowledge.
The authors declare no competing financial interests.\pagebreak{}\newpage{}


\pagebreak{}

\pagebreak{}

\pagebreak{}

\pagebreak{}\newpage{}

\begin{table}[H]
\caption{\label{tab:Pearson-correlation-coefficients}Pearson correlation coefficients
for different features related to monthly deployment.}

\begin{tabular}{|c|c|}
\hline 
Features & Pearson Correlation\tabularnewline
\hline 
\hline 
Mean IRR \textendash{} Monthly Deployment & 0.47\tabularnewline
\hline 
Mean Risk Adjusted IRR \textendash{} Monthly Deployment & 0.57\tabularnewline
\hline 
Exponential Utility Model \textendash{} Monthly Deployment & 0.62\tabularnewline
\hline 
Prospect Utility Model \textendash{} Monthly Deployment  & 0.85\tabularnewline
\hline 
\end{tabular}
\end{table}
\begin{landscape}

\begin{table}
\caption{\label{tab:Input-parameters}Input parameters and distributions used
in the study.}

\begin{tabular}{|c|c|c|c|c|c|c|c|}
\hline 
Input Parameter & Abbreviation & Unit & Distribution & \quad{}Min\quad{} & \quad{}Mode\quad{} & \quad{}Max\quad{} & Derived from\tabularnewline
\hline 
\hline 
Specific Investment  & $I_{0}$ & Euro/$\textrm{kW}_{\textrm{p}}$ & Normal & \multicolumn{3}{c|}{Mean: time series, standard deviation: 10\%} & \citep{EuPDResearch2016}\tabularnewline
\hline 
Size & $s$ & $\textrm{kW}_{\textrm{p}}$ & Uniform & >0 & 5 & 10 & Model assumption\tabularnewline
\hline 
Performance Ratio & $PR$ & \% & Beta & 75 & 84 & 90 & \citep{Reich2012}\tabularnewline
\hline 
Self-consumption Ratio & $SC$ & \% & Beta & 0 & 5 & 20 & \citep{Luthander2015}\tabularnewline
\hline 
Module Degradation & $d$ & \%/a & Beta & 0.0 & 0.5 & 2.0 & \citep{Jordan2011}\tabularnewline
\hline 
Roof Inclination & $\gamma$ & 1 & Beta & 0.25 & 0.98 & 1.00 & \citep{Mainzer2014}\tabularnewline
\hline 
Optimal Solar Potential & $H_{0}$ & kWh/m\texttwosuperior{}/a & Beta & 1141 & 1253 & 1403 & \citep{Huld2012}\tabularnewline
\hline 
O\&M Share & $c_{O\&M}$ & \% & Normal & \multicolumn{3}{c|}{Mean: 1.5, standard deviation: 0.15} & \citep{Hoppmann2014}\tabularnewline
\hline 
Retail Electricity Price & $e$ & Euro/kWh & Normal & \multicolumn{3}{c|}{Mean: time series, standard deviation: 5\%} & \citep{BDEW2016}\tabularnewline
\hline 
Lifetime & $T$ & years & Constant & \multicolumn{3}{c|}{20} & Model Assumption\tabularnewline
\hline 
\end{tabular}
\end{table}

\end{landscape}
\end{document}